\documentclass[11pt,a4paper]{article}

\usepackage[]{hyperref}
\usepackage{amsmath, amsthm, amssymb}
\usepackage{graphicx}

\def\ket#1{\left|#1\right>}
\def\bra#1{\langle#1\vert}
\def\exp#1{\langle#1\rangle}

\title{An argument for $\psi$-ontology \\ in terms of protective measurements}

\author{Shan Gao
\\Institute for the History of Natural Sciences, 
\\ Chinese Academy of Sciences, Beijing 100190, China. 
\\ E-mail:  \href{mailto:gaoshan@ihns.ac.cn}{gaoshan@ihns.ac.cn}.}

\begin{document}
\maketitle

%\vspace{2mm}
%\newpage
%\tableofcontents

\begin{abstract}\noindent 
The ontological model framework provides a rigorous approach to address the question of whether the quantum state is ontic or epistemic. When considering only conventional projective measurements, auxiliary assumptions are always needed to prove the reality of the quantum state in the framework. For example, the Pusey-Barrett-Rudolph theorem is based on an additional preparation independence assumption. In this paper, we give a new proof of $\psi$-ontology in terms of protective measurements in the ontological model framework. The proof does not rely on auxiliary assumptions, and also applies to deterministic theories such as the de Broglie-Bohm theory. In addition, we give a simpler argument for $\psi$-ontology beyond the framework, which is only based on protective measurements and a weaker criterion of reality. The argument may be also appealing for those people who favor an anti-realist view of quantum mechanics.
\end{abstract}

\vspace{6mm}

\section{Introduction}

The nature of the quantum state has been a hot topic of debate since the early days of quantum mechanics. A long-standing question is whether the quantum state assigned to a single system represents the physical state of the system or a state of (incomplete) knowledge about the physical state of the system (Einstein, Podolsky and Rosen, 1935). In recent years, the framework of ontological models provides a rigorous approach to address this question by formalizing the distinction between these two views, which are referred to as $\psi$-ontic and $\psi$-epistemic, respectively (Spekkens, 2005; Harrigan and Spekkens, 2010). Several theorems have also been proved to establish the $\psi$-ontic view within the framework (Pusey, Barrett and Rudolph, 2012; Colbeck and Renner, 2012; Hardy, 2013; Patra, Pironio and Massar, 2013)\footnote{For a comprehensive review of these $\psi$-ontology theorems and related work see Leifer (2014).}. However, on the one hand, the ontological model framework is not very general, and on the other hand, auxiliary assumptions are always required to prove these $\psi$-ontology theorems, e.g. the preparation independence assumption for the Pusey-Barrett-Rudolph theorem. It thus seems to be impossible to completely rule out $\psi$-epistemic models for quantum mechanics without auxiliary assumptions. Indeed, by removing the assumptions of these $\psi$-ontology theorems, explicit $\psi$-epistemic models can be constructed to reproduce the statistics of quantum measurements in Hilbert spaces of any dimension (Lewis et al, 2012; Aaronson et al, 2013). 

In this paper, we will give a new argument for $\psi$-ontology in terms of protective measurements, first in the ontological model framework and then beyond the framework. Protective measurements are distinct from projective measurements in that a protective measurement can directly obtain the expectation value of the measured observable in the measured state with certainty (Aharonov and Vaidman, 1993; Aharonov, Anandan and Vaidman, 1993), while a projective measurement can only obtain one of the eigenvalues of the measured observable with certain probability in accordance with the Born rule. As a consequence, the existence of protective measurements will extend the ontological model framework, and provide more resources for proving the reality of the quantum state. 

The plan of this paper is as follows. In Section 2, we give a concise introduction to protective measurements. It is shown that the appearance of expectation value as a measurement result is quite natural when the measured state is not changed during the measurement as for protective measurements. In Section 3, we present a new, rigorous proof of $\psi$-ontology in the extended ontological model framework which includes protective measurements. The proof needs not rely on auxiliary assumptions, and also applies to deterministic theories such as the de Broglie-Bohm theory. In Section 4, we further improve the ontological model framework by replacing one of its fundamental assumptions with a more reasonable one. We argue that although the proofs of existing $\psi$-ontology theorems cannot go through under this new assumption, our proof is still valid. In Section 5, we replace the ontological model framework with a weaker criterion of reality, which is arguably an improved version of the Einstein-Podolsky-Rosen criterion of reality, and give a simpler argument for $\psi$-ontology based on protective measurements and this criterion of reality. Conclusions are given in the last section.

\section{Protective measurements}

The existing $\psi$-ontology theorems and $\psi$-epistemic models are both based on an analysis of conventional projective measurements\footnote{It is worth emphasizing that the existing $\psi$-epistemic models reproduce only the statistics of conventional projective  measurements, not yet the outcomes of protective measurements (Spekkens, 2007; Lewis et al, 2012; Aaronson et al, 2013).}. However, there are in fact other types of quantum measurements, one of which is the important but seemingly less-known protective measurements (Aharonov and Vaidman, 1993; Aharonov, Anandan and Vaidman, 1993). During a protective measurement, the measured state is protected by an appropriate mechanism such as via the quantum Zeno effect, so that it neither changes nor becomes entangled with the state of the measuring device. In this way, such protective measurements can measure the expectation values of observables on a single quantum system, even if the system is initially not in an eigenstate of the measured observable, and the quantum state of the system can also be measured as expectation values of a sufficient number of observables.

By a projective measurement on a single quantum system, one obtains one of the eigenvalues of the measured observable, and the expectation value of the observable can only be obtained as the statistical average of eigenvalues for an ensemble of identically prepared systems. Thus it seems surprising that a protective measurement can obtain the expectation value of the measured observable directly from a single quantum system. In fact, the appearance of expectation value as a measurement result is quite natural when the measured state is not changed during the measurement as for protective measurements (Aharonov, Anandan and Vaidman, 1993). In this case, the evolution of the combining state is

\begin{equation}
\ket{\psi(0)}\ket{\phi(0)} \rightarrow \ket{\psi(t)}\ket{\phi(t)}, t>0
\end{equation}

\noindent where $\ket{\psi}$ denotes the state of the measured system and $\ket{\phi}$ the state of the measuring device, and $\ket{\psi(t)}$ is the same as $\ket{\psi(0)}$ up to a phase factor during the measurement interval $[0,\tau]$. The interaction Hamiltonian is given by $H_I = g(t)PA$,  where $A$ is the measured observable, $P$ is the conjugate momentum of the pointer variable $X$ of the device, and the time-dependent coupling strength $g(t)$ is a smooth function normalized to $\int dt g(t)=1$ during the measurement interval $\tau$, and $g(0)=g(\tau)=0$. Then by Ehrenfest's theorem we have

\begin{equation}
{d \over dt}\bra{\psi(t)\phi(t)}X\ket{\psi(t)\phi(t)} = g(t)\bra{\psi(0)}A\ket{\psi(0)},
\end{equation}

\noindent which further leads to

\begin{equation}
\bra{\phi(\tau)}X\ket{\phi(\tau)}- \bra{\phi(0)}X\ket{\phi(0)}= \bra{\psi(0)}A\ket{\psi(0)}.
\end{equation}

\noindent This means that the shift of the center of the pointer of the device gives the expectation value of the measured observable in the measured state. This analysis also shows that a protective measurement obtaining an expectation value is independent of the protection procedure.

That the quantum state of a single system can be measured by protective measurements can also be illustrated with a specific example (Aharonov and Vaidman, 1993). Consider a quantum system in a discrete nondegenerate energy eigenstate $\psi(x)$. In this case, the measured system itself supplies the protection of the state due to energy conservation and no artificial protection is needed. We take the measured observable $A_n$ to be (normalized) projection operators on small spatial regions $V_n$  having volume $v_n$:

\begin{equation}
A_n= 
\begin{cases} 
{1\over{v_n}},& \text{if $x \in V_n$,}
\\
0,&\text{if $x \not\in V_n$.} 
\end{cases}
\end{equation}

\noindent An adiabatic measurement of $A_n$ then yields 

\begin{equation}
\exp{A_n} = {1\over {v_n}} \int_{V_n}|\psi(x)|^2 dv,
\end{equation}

\noindent  which is the average of the density $\rho(x) = |\psi(x)|^2$ over the small region $V_n$.  Similarly, we can adiabatically measure another observable $B_n ={\hbar \over{2mi}} (A_n\nabla + \nabla A_n)$. The measurement yields 

\begin{equation}
\exp{B_n} ={1\over {v_n}} \int_{V_n}{\hbar \over{2mi}}(\Psi^* \nabla \Psi - \Psi  \nabla \Psi^* )dv = {1\over {v_n}} \int_{V_n}j(x)dv.
\end{equation}

\noindent This is the average value of the flux density $j(x)$ in the region $V_n$. Then when $v_n \rightarrow 0$ and after performing measurements in sufficiently many regions $V_n$ we can measure $\rho(x)$ and $j(x)$ everywhere in space. Since the quantum state $\psi(x,t)$ can be uniquely expressed by $\rho(x,t)$ and $j(x,t)$ (except for an overall phase factor), the above protective measurements can obtain the quantum state of the measured system.

\section{My argument}

Since the quantum state can be measured from a single system by a series of protective measurements, it seems natural to assume that the quantum state refers directly to the physical state of the system. Several authors, including the discoverers of protective measurements, have given similar arguments supporting this implication of protective measurements for the ontological status of the quantum state (Aharonov and Vaidman 1993; Aharonov, Anandan and Vaidman 1993; Anandan 1993; Dickson 1995; Gao 2013, 2014a; Hetzroni and Rohrlich 2014). However, these analyses are not very rigorous and also subject to some objections (Unruh 1994; Dass and Qureshi 1999; Schlosshauer and Claringbold 2014)\footnote{See Gao (2014b, 2016) for a brief review of and answers to these objections.}. It is still debatable whether protective measurements imply the reality of the quantum state. In the following, we will give a new, rigorous argument for $\psi$-ontology in terms of protective measurements in the ontological model framework.

The ontological model framework is based on two fundamental assumptions (Harrigan and Spekkens, 2010; Pusey, Barrett and Rudolph, 2012). The first one is that if a quantum system is prepared such that quantum theory assigns a pure state, then after preparation the system has a well-defined set of physical properties, which is usually represented by a mathematical object, $\lambda$. This assumption is necessary for the analysis of the ontological status of the quantum state, since if such physical properties don't exist, it will be meaningless to ask whether or not the quantum state describes them. The second assumption is that when a measurement is performed, the behaviour of the measuring device is only determined by the complete physical state of the system, along with the physical properties of the measuring device. For a projective measurement $M$, this assumption means that the physical state or ontic state $\lambda$ of a system determines the probability $p(k|\lambda,M)$ of different outcomes $k$ for the measurement $M$ on the system. While for a protective measurement, this assumption will mean that the ontic state $\lambda$ of a system determines the definite result of the protective measurement on the system. In this way, the ontological model framework is enlarged by including protective measurements, and it also applies to theories in which the Born probability is not determined by the ontic state of the measured system, such as the de Broglie-Bohm theory (see also Drezet, 2015).

Based on these assumptions, we can give an rigorous argument for $\psi$-ontology in terms of protective measurements. 
We first use the proof strategy of existing $\psi$-ontology theorems, namely first assuming that two different quantum states are compatible with the same ontic state, and then proving the consequences of this assumption are inconsistent with the predictions of quantum mechanics. The argument is as follows. For two different quantum states such as two nonorthogonal states, select an observable whose expectation values in these two states are different. For example, consider a spin half particle. The two nonorthogonal states are $\ket{0}$ and  $\ket{+}$, where $\ket{+}= {1 \over \sqrt{2}}(\ket{0} + \ket{1})$, and $\ket{0}, \ket{1}$ are eigenstates of spin in the z-direction. As Aharonov, Anandan and Vaidman (1993) showed, a spin state can be protected by a magnetic field in the direction of the spin. Let $B_0, B_+$ be protecting fields for the states $\ket{0}$, $\ket{+}$, respectively, and let the measured observable  be $P_0=\ket{0}\bra{0}$. Then the protective measurements of this observable on these two nonorthogonal states yield results 1 and 1/2, respectively. Although these two nonorthogonal states need different protection procedures, the protective measurements of the observable on the two (protected) states are the same, and the results of the measurements are different with certainty. If there exists a probability $p>0$ that these two (protected) quantum states correspond to the same ontic state $\lambda$, then according to the assumptions of the ontological model framework, the results of the protective measurements of the observable on these two states will be the same with probability not smaller than $p$. This leads to a contradiction. Therefore, two (protected) quantum states correspond to different ontic states\footnote{This result is not surprising, since two (protected) quantum states of a single system can be distinguished with certainty by protective measurements.}. By assuming that whether an unprotected state or a corresponding protected state is prepared, the probability distribution of the ontic state $\lambda$ is the same, which is an inference of the preparation noncontextuality assumption (Spekkens, 2005; Leifer, 2014), we can further reach the conclusion that two (unprotected) quantum states also correspond to distinct ontic states. In other words, the quantum state represents the physical state of a single  system.

A similar argument can also be given in terms of realistic protective measurements. A realistic protective measurement cannot be performed on a single quantum system with absolute certainty. For a realistic protective measurement of an observable $A$, there is always a small probability to obtain an outcome different from $\exp{A}$. In this case, according to the assumptions of the ontological model framework, the probability of different outcomes will be determined by the ontic state of the measuring device and the realistic measuring condition such as the measuring time, as well as by the ontic state of the measured system\footnote{Similarly, the probability of different outcomes of a realistic projective measurement will be also determined by the ontic state of the measuring device and the measuring time, as well as by the ontic state of the measured system. As we will see later, the existing $\psi$-ontology theorems will be difficult or even impossible to prove for realistic projective measurements.}. Now consider two  (protected) quantum states, and select an observable whose expectation values in these two states are different. Then we can perform the same realistic protective measurements of the observable on these two states. The overlap of the probability distributions of the results of these two measurements can be arbitrarily close to zero when the realistic condition approaches the ideal condition (In the limit, each probability distribution will be a Dirac $\delta-$function localized in the expectation value of the measured observable in the measured state, and it will be determined only by the ontic state of the measured system). If there exists a non-zero probability $p$ that these two quantum states correspond to the same ontic state $\lambda$, then since the same $\lambda$ yields the same probability distribution of measurement results under the same measuring condition according to the ontological model assumptions, the overlap of the probability distributions of the results of protective measurements of the above observable on these two states will be not smaller than $p$. Since $p>0$ is a determinate number, this leads to a contradiction\footnote{Note that it is indeed true that for any given realistic condition one can always assume that there exists some probability $p$ that the two measured quantum states correspond to the same ontic state $\lambda$. However, the point is that if the unitary dynamics of quantum mechanics is valid, the realistic condition can always approach the ideal condition arbitrarily closely, and thus the probability $p$ must be arbitrarily close to zero, which means that any $\psi$-epistemic model with finite overlap probability $p$ is untenable. Certainly, our argument will be invalid if quantum mechanics breaks down when reaching certain realistic condition.}. Therefore, two (protected) quantum states correspond to different ontic states, and so do two (unprotected) quantum states by the preparation noncontextuality assumption.

The above argument, like the existing $\psi$-ontology theorems, also needs to be based on an auxiliary assumption, the preparation noncontextuality assumption this time\footnote{Since the above argument only considers single quantum systems and makes no appeal to entanglement, it avoids the preparation independence assumption for multiple systems used by the Pusey-Barrett-Rudolph theorem (Pusey, Barrett and Rudolph, 2012).}. However, the argument can be further improved to avoid this auxiliary assumption. The key is to notice that the result of a protective measurement depends only on the measured observable and the ontic state $\lambda$ of the measured system. If the result is also determined by other factors such as the ontic state of the measuring device or the protection setting, then the result may be different for the same measured observable and quantum state. This contradicts the predictions of quantum mechanics, according to which the result of a protective measurement is always the expectation value of the measured observable in the measured quantum state. Now consider  two  (unprotected) quantum states, and select an observable whose expectation values in these two states are different. The results of the protective measurements of the observable on these two states are different with certainty. If there exists a probability $p>0$ that these two quantum states correspond to the same ontic state $\lambda$, then according to the above analysis, the results of the protective measurements of the observable on these two states will be the same with probability not smaller than $p$. This leads to a contradiction. Therefore, two distinct quantum states correspond to different ontic states.

We can also give a direct argument for $\psi$-ontology in terms of protective measurements, which is not based on auxiliary assumptions either. As argued above, the result of a protective measurement is determined only by the measured observable and the ontic state $\lambda$ of the measured system. Since the measured observable also refers to the measured system, this further means that the result of a protective measurement, namely the expectation value of the measured observable in the measured quantum state, is determined only by the realistic properties of the measured system. Therefore, the expectation value of the measured observable is also a realistic property of the measured system. In other words, the expectation value of an observable is a realistic property of a single quantum system. Since a quantum state can be constructed from the expectation values of a sufficient number of observables, the quantum state is also real.

\section{With more strength}

The above arguments for $\psi$-ontology, like the existing $\psi$-ontology theorems, are also based on the second assumption of the ontological model framework, which says that for a protective measurement the ontic state of a physical system immediately before the measurement determines the outcome of the measurement, whether the ontic state changes or not during the measurement. However, this is certainly a simplified assumption. A more reasonable assumption is that the ontic state of a physical system may evolve in a certain way during a protective measurement, and the measurement outcome is determined not only by the initial ontic state but also by the total evolution of the ontic state during the measurement. Similarly, it is the total evolution of the ontic state of a system during a projective measurement determines the probability of different outcomes for the measurement. Certainly, if the measuring interval is extremely short and the change of the ontic state is continuous, then the ontic state will be almost unchanged during the measurement, and thus the original simplified assumption will be still valid. However, if the change of the ontic state is not continuous but discontinuous, then even during an arbitrarily short time interval the ontic state may change greatly, and thus the original, simplified assumption will be invalid. 

Unfortunately, the proofs of existing $\psi$-ontology theorems such as the Pusey-Barrett-Rudolph theorem will not go through under this more reasonable assumption. The reason is that under this assumption, even if two nonorthogonal states correspond to the same ontic state initially, they may correspond to different evolution of the ontic state, which may then lead to different probabilities of measurement outcomes. Then the proofs of the $\psi$-ontology theorems by reduction to absurdity cannot go through. In a similar way, the above arguments in terms of protective measurements, which use the proof strategy of existing $\psi$-ontology theorems, will be also invalid under this more reasonable assumption. 

However, the direct argument for $\psi$-ontology in terms of protective measurements can still go through under the new assumption. First, according to this assumption, the evolution of the ontic state of a physical system during a protective measurement determines the result of the protective measurement, namely the expectation value of the measured observable in the measured quantum state. Next,  since the quantum state of the system keeps unchanged, the evolution of the ontic state of the system is still compatible with the quantum state. This means that even when the system being in the quantum state is not measured, its ontic state may also evolve in this way and such evolution is then a realistic property of the system. Therefore, the expectation value of the measured observable is determined by a realistic property of the measured system, and it is also a realistic property of the system. Then similar to the direct argument given in the last section, we can also prove the reality of the quantum state.

\section{A weaker criterion of reality}

The first fundamental assumption of the ontological model framework is that a quantum system after preparation has a well-defined set of realistic properties (Harrigan and Spekkens, 2010; Pusey, Barrett and Rudolph, 2012). The existing $\psi$-ontology theorems, as well as the above arguments in terms of protective measurements, are all based on this realistic assumption. If one drops this assumption as anti-realists would like to do, then one can still restore the (non-realist) $\psi$-epistemic view or assume another non-realist view. In this section, we will give an argument for the reality of the quantum state based not on this realistic assumption but on a weaker criterion of reality. The analysis is beyond the ontological model framework.

The suggested criterion of reality is: if a certain observation of a physical system obtains a definite outcome, which is determined by a mathematical quantity assigned to the system by a theory, and the quantity does not change during the observation, then the system has a definite, realistic property represented by that quantity according to the theory\footnote{This criterion of reality improves the Einstein-Podolsky-Rosen criterion of reality in that it avoids the requirement of ``without in any way disturbing a system", which is difficult or even impossible to justify (as we don't know the underlying ontic state of the measured system and its possible dynamics during a measurement before the analysis using the criterion of reality), and replaces it with a more reasonable condition that the measured quantity does not change during the measurement, whose validity can be precisely determined by a realist theory.}. This criterion of reality provides a definite link from the mathematical quantities in a realistic theory to the realistic properties of a physical system via experience or measurements. By using this criterion of reality to analyze the ontological content of a theory such as the ontological status of the quantum state in quantum mechanics, we need not care about the underlying ontic state of a physical system and its possible dynamics during a measurement. Here is the analysis. A protective measurement on a physical system yields a definite outcome, the expectation value of the measured observable in the measured quantum state. This outcome is determined by the quantum state, a mathematical quantity assigned to the measured system, and the quantum state also keeps unchanged during the measurement. Therefore, according to the suggested criterion of reality, the system has a definite, realistic property represented by the quantum state. This proves the reality of the quantum state.

Since the above criterion of reality does not necessarily require that a quantum system have realistic properties, it is weaker than the realistic assumption of the ontological model framework. Even though some people refuse to attribute realistic properties to quantum systems, they may well accept this criterion of reality. On the one hand, this criterion of reality perfectly applies to classical mechanics, and one can use it to get the anticipant ontological content of the theory. On the other hand, people usually think that this criterion of reality cannot be applied to quantum mechanics in general (though it can certainly apply to the measurements of eigenstates of an observable), and thus it does not influence the non-realist views of the theory. However, the existence of protective measurements must be a surprise for these people. It will be interesting to see whether some anti-realists will reject this criterion of reality based on the existence of protective measurements.

Certainly, one can also restore the (non-realist) $\psi$-epistemic view by rejecting the above criterion of reality. However, there is a good reason why this is not a good choice. It is arguably that a reasonable, universal criterion of reality, which may provide a plausible link between theory and reality via experience, is useful or even necessary for realist theories. The criterion of reality is not necessarily complete, being able to derive all ontological content of a theory, which seems to be an impossible task. However, we can at least derive the basic ontological content of a realist theory by using this criterion of reality. If one admits the usefulness and universality of such a criterion of reality, then the similarity between classical measurements and protective measurements will require that if one assumes a realist view of classical mechanics, admitting the ontological content of the theory derived from the suggested criterion of reality, then one must also admit the ontological content of quantum mechanics derived from this criterion of reality, such as the reality of the quantum state. The essential point is not that the suggested criterion of reality must be true, but that if we accept the usefulness and universality of such a  criterion of reality and apply it to classical mechanics and macroscopic objects to derive the anticipant classical ontology, we should also apply it to quantum mechanics and microscopic objects to derive the unexpected quantum ontology, no matter how strange it is. Otherwise we will have to divide the world into a quantum one and a classical one artificially, and we will not have a unified world view as a result.

\section{Conclusion}

In this paper, we present a new, rigorous proof of $\psi$-ontology in terms of protective measurements. The analysis improves the previous work in several aspects. First of all, we extend the ontological model framework by introducing protective measurements, and prove the reality of the quantum state in this framework without relying on auxiliary assumptions. This provides a stronger $\psi$-ontology theorem. Moreover, the extended ontological model framework also applies to deterministic theories such as the de Broglie-Bohm theory, so does our proof of $\psi$-ontology in the framework. Next, we further improve the ontological model framework by replacing one of its fundamental assumptions with a more reasonable one. We argue that although the proofs of existing $\psi$-ontology theorems cannot go through under the new assumption, our proof is still valid. Lastly, we replace the ontological model framework with a weaker criterion of reality, and give a simpler argument for $\psi$-ontology based on protective measurements and this criterion of reality. The argument may be also appealing for some people who favor an anti-realist view of quantum mechanics.

\section*{Acknowledgments}
I thank Robert Griffiths, Matthew Leifer, Matthew Pusey, Max Schlosshauer and Ken Wharton  for helpful discussions at the First iWorkshop on the Meaning of the Wave Function organized by International Journal of Quantum Foundations. I am also grateful to two anonymous referees and the co-editor of this journal, Wayne Myrvold, for their insightful comments, constructive criticisms and helpful suggestions. This work is partly supported by the Top Priorities Program of the Institute for the History of Natural Sciences, Chinese Academy of Sciences under Grant No. Y45001209G.

\section*{References}
\renewcommand{\theenumi}{\arabic{enumi}}
\renewcommand{\labelenumi}{[\theenumi]}
\begin{enumerate}

\item Aaronson S., Bouland A., Chua L. and Lowther G. (2013). $\psi$-epistemic theories: The role of symmetry. Phys. Rev. A 88, 032111.
\item Aharonov, Y., Anandan, J. and Vaidman, L. (1993). Meaning of the wave function. Phys. Rev. A 47, 4616.
\item Aharonov, Y. and Vaidman, L. (1993). Measurement of the Schr\"{o}dinger wave of a single particle. Phys. Lett. A 178, 38.
\item Anandan,	J.  (1993). Protective measurement and quantum reality. Found. Phys. Lett., 6, 503-532.
\item Colbeck, R., and Renner, R. (2012). Is a system's wave function in one-to-one correspondence with its elements of reality? Phys. Rev. Lett., 108, 150402.
\item Dass, N. D. H. and Qureshi, T. (1999). Critique of protective measurements. Phys. Rev. A 59, 2590.
\item Dickson, M. (1995). An empirical reply to empiricism: protective measurement opens the door for quantum realism. Philosophy of Science 62, 122.
\item Drezet, A. (2015). The PBR theorem seen from the eyes of a Bohmian. International Journal of Quantum Foundations, 1, 25-43.
\item Einstein, A., Podolsky, B. and Rosen, N. (1935). Can quantum-mechanical description of physical reality be considered complete? Phys. Rev. 47, 777.
\item Gao, S. (2013). On Uffink's criticism of protective measurements, Studies in History and Philosophy of Modern Physics, 44, 513-518.
\item Gao, S. (2014a). Reality and meaning of the wave function. In S. Gao (eds.). Protective Measurements and Quantum Reality: Toward a New Understanding of Quantum Mechanics. Cambridge: Cambridge University Press. pp.211-229.
\item Gao, S. (2014b). Protective measurement: An introduction. In S. Gao (eds.). 
Protective Measurements and Quantum Reality: Toward a New Understanding of Quantum Mechanics. Cambridge: Cambridge University Press. pp.1-12.
\item Gao, S. (2016). Meaning of the Wave Function: In Search of the Ontology of Quantum Mechanics. Cambridge: Cambridge University Press (forthcoming).
\item Hardy, L. (2013). Are quantum states real? International Journal of Modern Physics B 27, 1345012.
\item Harrigan, N. and Spekkens, R. (2010). Einstein, incompleteness, and the epistemic view of quantum states. Found. Phys. 40, 125-157.
\item Hetzroni, G. and Rohrlich, D. (2014). Protective measurements and the PBR theorem. In S. Gao (eds.). Protective Measurements and Quantum Reality: Toward a New Understanding of Quantum Mechanics. Cambridge: Cambridge University Press. pp.135-144.
\item Leifer, M. S. (2014). Is the quantum state real? An extended review of $\psi$-ontology theorems, Quanta 3, 67-155.
\item Lewis, P. G., Jennings, D., Barrett, J., and Rudolph, T.  (2012). Distinct quantum states can be compatible with a single state of reality. Phys. Rev. Lett. 109, 150404.
\item Patra, M. K., Pironio,  S. and Massar, S. (2013). No-go theorems for $\psi$-epistemic models based on a continuity assumption. Phys. Rev. Lett., 111, 090402.
\item Pusey, M., Barrett, J. and Rudolph, T. (2012). On the reality of the quantum state. Nature Phys. 8, 475-478.
\item Schlosshauer, M. and Claringbold, T. V. B. (2014). Entanglement, scaling, and the meaning of the wave function in protective measurement. In S. Gao (eds.). Protective Measurements and Quantum Reality: Toward a New Understanding of Quantum Mechanics. Cambridge: Cambridge University Press. pp.180-194. 
\item Spekkens, R. W. (2005). Contextuality for preparations, transformations, and unsharp measurements. Phys. Rev. A 71, 052108. 
\item Spekkens, R. W. (2007). Evidence for the epistemic view of quantum states: A toy theory. Phys. Rev. A 75, 032110.
\item Unruh, W. G. (1994). Reality and measurement of the wave function. Phys. Rev. A 50, 882.

\end{enumerate}
\end{document}